\documentclass[prb,twocolumn,aps,superscriptaddress,showpacs,floatfix]{revtex4}
\usepackage{amsmath}
\usepackage{graphicx}
\usepackage{color}
\usepackage{tikz}
\begin{document}
\title{Decoherence of Majorana qubits caused by particle-hole conjugation asymmetry}
\author{R.S. Akzyanov}
\affiliation{Moscow Institute of Physics and Technology, Dolgoprudny,
Moscow Region, 141700 Russia}
\affiliation{Institute for Theoretical and Applied Electrodynamics, Russian
Academy of Sciences, Moscow, 125412 Russia}
\affiliation{Dukhov Research Institute of Automatics, Moscow, 127055 Russia}

\begin{abstract}
We investigate effects of the particle-hole asymmetry on the properties of Majorana qubtis. Particle-hole asymmetric pertrubation shifts Majorana zero mode from zero-energy. This type of asymmetry can act as a source of decoherence in Majorana qubits. We find that particle-hole asymmetric pertrubation causes phase-shifts and bit-flips decoherence processes in Majorana qubits. However, these sources of decoherence are negligible in uniform system with spatially separated Majorana zero modes.
\end{abstract}


\maketitle
\section{Introduction}
Topological quantum computations (TQC) is promising scheme for the fault-tolerant quantum codes. Key advantage of such computations is their robustness to the local sources of decoherence. Realization of such codes requires non-abelian anyons as a building blocks. Braiding of anyons can be used for realization of quantum gates.~\cite{kitaev_tqc,tqc}

 One of the simplest realization of non-abelian anyons in condensed matter physics is Majorana zero modes (MZM)~\cite{Alicea_review,Leijnse_review,Beenakker_review,Franz_review}. Such modes can be realized at the ends of superconducting nanowires with spin-orbit coupling~\cite{Stanescu_Tewari} and topological insulator nanowires~\cite{Franz_nanowire}; at the vortices in $p$-wave superconductors~\cite{Ivanov_Phys.Rev.Lett._2001, read_green} and topological insulator - superconductor heterostructures~\cite{fu_kane_device, me, me_edge}; as a Moore-Read state in fractional quantum Hall effect~\cite{moore_read}. Recent experiments show signatures of MZM in the superconducting topological nanowires~\cite{exp1} and topological insulator - superconductor structures~\cite{chinese}.

 MZM can be used for constructing parity qubits that are robust against the decoherence~\cite{Beenakker_review,akhmerov_qubits} and to disorder~\cite{akhmerov_disorder}. In this type of qubits occupation number of fermion level constructed from two MZM defines qubit space.  However, robustness of parity qubits can be lifted by a MZM coupling~\cite{Beenakker_review,maj_tunneling}.

 Previously, decoherence of Majorana qubits have been studied extensively~\cite{d1,d2,d3}. It has been argued that tunneling of electrons from the environment to the topological superconductor can destroy the coherence of the parity qubit causing bit-flip errors by changing parity~\cite{d1}. However, recent experiments show that parity lifetime can reach a large values~\cite{de1,de2}.

Topological states of matter can be classified by their global symmetries: time-reversal and particle-hole conjugation (PHC) symmetries and their combination, chiral symmetry. For example, system with unitary PHC $\Xi^2=+1$ and broken time-reversal symmetry $T^2=0$ is in class $D$ supports MZM at a point defects~\cite{kane_defects}. If PHC symmetry is broken then symmetry of such a system reduces from the class $D$ to the class $A$. In that class topological index is trivial for point defects, hence, MZM cannot appear~\cite{kane_defects}.

Indeed, MZM is symmetry protected state that is governed by the PHC symmetry. Usually, this symmetry is hard to violate in superconductors. However, even small PHC asymmetry can lead to the experimentally observable features, such as vortex charge\cite{vortex_charge}. Nevertheless, possible effects of violation of PHC symmetry and its effects on decoherence in Majorana qubits have to be studied to find out optimal parameters for topological quantum computations.

Aim of our work is to study limitations of MZM based quantum computations in presence of PHC asymmetric pertrubation. First, we study general effects of the PHC asymmetric pertrubation on the MZM. Second, we study decoherence in the MZM based parity qubits. Finally, we study possible origins of such PHC asymmetry in condensed matter systems.
\section{Perturbation theory}
We start from the general superconducting Hamiltonian that has unitary PHC symmetry:
\begin{eqnarray}\label{phc_symm}
\Xi H \Xi = -H\\
\Xi^2 = +1
\end{eqnarray}
where $H$ is the Hamiltonian, $\Xi$ is the particle-hole conjugation operator. We suppose that Hamiltonian has MZM solution
\begin{eqnarray}
H \gamma=0\\
\Xi\gamma=\gamma
\end{eqnarray}
This MZM is robust to any kind of pertrubations that do not close the gap in the spectrum. Only existence of another MZM can shift zero-energy degeneracy and split zero energy. This splitting is exponentially small if the distance between two MZM is large~\cite{maj_tunneling}. However, previous analysis do not consider pertrubations that violate PHC symmetry. Under this type of petrubations MZM is no longer robust.

In general, pertrubation can be written as the sum of two parts:
\begin{eqnarray}
V=V_s+V_{as}\nonumber\\
\Xi V_s\Xi= -V_s \nonumber\\
\Xi V_{as}\Xi= +V_{as}
\end{eqnarray}
where $V_s$ is the pertrubation that conserve PHC symmetry, while $V_{as}$ violates this type of symmetry. It can be shown that if $\gamma=\Xi \gamma$ is MZM then
$
\langle \gamma|V_s|\gamma \rangle = -\langle \gamma|V_s|\gamma \rangle=0
$. So, only PHC asymmetric $V_{as}$ can shift single MZM form zero energy. In this case energy correction to the zero-energy of MZM can be calculated as
\begin{equation}
\delta E = \langle \gamma |V_{as}| \gamma \rangle
\end{equation}
Corrections to the wavefunction can be calculated as follows
\begin{eqnarray}\label{corr}
\delta \gamma = \delta_s \gamma+\delta_{as} \gamma \\
\delta_s \gamma = \sum \limits_{n\neq0}\frac{\langle \psi_n |V_s|\gamma\rangle }{-E_n} |\psi_n\rangle = +\Xi\delta_s \gamma \\
\delta_{as} \gamma = \sum \limits_{n\neq0}\frac{\langle \psi_n |V_{as}|\gamma\rangle }{-E_n} |\psi_n\rangle = -\Xi\delta_{as} \gamma
\end{eqnarray}
where $\delta_s \gamma$ is the correction of the wave function caused by the PHC symmetric pertrubation $V_s$, while $\delta_{as} \gamma$ is the correction of the wave function caused by the PHC asymmetric pertrubation $V_{as}$; $E_n$ is the energy of the level with wavefunction $\psi_n$.

It can be shown from Eq.~\ref{corr} that $\gamma+\delta_s \gamma=\Xi (\gamma+\delta_s \gamma)$ and $i\delta_{as} \gamma=\Xi i\delta_{as} \gamma$ obey condition for MZM. So, perturbed state $\gamma +\delta\gamma$ can be represented as a superposition of two terms, each of them can be formally written as MZM: $\gamma+\delta_s \gamma$ and $i\delta_{as} \gamma$. How we can interpret this result? Similar situation arises when we consider splitting caused by tunneling of two MZM~\cite{maj_tunneling}: Dirac state with non-zero energy is written as superposition of two separate MZM $\gamma_1+i\gamma_2$. In our case, we have one MZM $\gamma$. Correction to the MZM state $\delta_{as} \gamma$ can be interpreted as a 'fake' MZM that couples to the initial MZM $\gamma$. By 'fake' we means that $\delta_{as} \gamma$ do not correspond to any physical state alone. However, this analogy gives us a tool to analyze decoherence effects of PHC asymmetric pertrubation in topological quantum computations.
\section{PHC asymmetry on topological quantum computation}
In this section we investigate stability of TQC versus PHC asymmetry pertrubation.

 We start with the case without PHC asymmetric pertrubation $V_{as}=0$. Two separate MZM $\hat{\gamma_1}$ with wavefunction $\gamma_1$ and $\hat{\gamma_2}$ with wavefunction $\gamma_2$ define one fermion level which can be empty $|0\rangle$ or full $|1\rangle$, ladder operators acts as follows $(\hat{\gamma_1}-i\hat{\gamma_2})/\sqrt{2}|1\rangle=|0\rangle$ and $(\hat{\gamma_1}+i\hat{\gamma_2})/\sqrt{2}|0\rangle=|1\rangle$.  However, qubit constructed by two MZM cannot be rotated since parity is conserved. Four MZM are needed to construct a qubit~\cite{Beenakker_review}. Without loss of generality we can assume that number of electrons is even. Then, general state of the qubit can be written as superposition of two states $|0\rangle_1 |0\rangle_2$ and $|1\rangle_1 |1\rangle_2$ because they have the same parity
\begin{eqnarray}
\Psi = \alpha|0\rangle_1 |0\rangle_2 + \beta |1\rangle_1 |1\rangle_2, \qquad |\alpha|^2+|\beta|^2=1
\end{eqnarray}
where $|.\rangle_1$ and $|.\rangle_2$ correspond to the first and the second fermion levels.
This parity qubit is protected from the phase-shift errors, since energy of both states is zero and the qubit do not rotate $\Psi(t)=\Psi(0)$. Also, parity qubit is protected against bit-flip errors that cause transition $|0\rangle \rightarrow |1\rangle$ because this process is forbidden by the conserved parity. Thus, parity qubit is topologically protected by PHC symmetry. However, this protection can be lifted if we consider tunneling between MZM~\cite{Beenakker_review,maj_tunneling}. By analogy between MZM coupling and PHC asymmetric perturbation we can conclude that similar effects can be observed in the last case $V_{as} \neq 0$.

 First, we study phase-shift error, that caused by dephasing between rotating qubits. Under PHC asymmetric perturbation MZM acquire non-zero energy. Thus, states with empty fermion level $|0\rangle$ which corresponds to the state $\psi_-=(\gamma_1-i\gamma_2)/\sqrt{2}$ and full fermion level $|1\rangle$ which corresponds to the state $\psi_+=(\gamma_1+i\gamma_2)/\sqrt{2}$ start rotating
 \begin{eqnarray}
 |0\rangle_t = e^{-iE_- t} |0\rangle_{t=0} \nonumber\\
 |1\rangle_t = e^{-iE_+ t} |1\rangle_{t=0}\\
 \end{eqnarray}
 where $E_+=\langle \psi_+|V_{as}|\psi_+\rangle \simeq E_-\simeq E$. Energy $E$ can be calculated as follows
 \begin{eqnarray}\label{rotating}
 E= (\langle \gamma_1|V_{as}|\gamma_1\rangle+\langle \gamma_2|V_{as}|\gamma_2\rangle)/2
 \end{eqnarray}
 We omit factors $\langle \gamma_2|V_{as}|\gamma_1 \rangle$ and $\langle \gamma_1|V_{as}|\gamma_2 \rangle$ since this matrix elements are exponentially small if distance between MZM is much larger than a coherence length. States with empty and full fermion level has the same energy. So, qubit evolves in time by the law
 \begin{eqnarray}
\Psi_t = e^{-it(E_1+E_2)}\Psi_{t=0}
\end{eqnarray}
and rotating with period
\begin{eqnarray}\label{T}
 T=2\pi/(E_1+E_2)
\end{eqnarray}
where $E_{1,2}$ is defined by Eq.~\ref{rotating} and belongs to the two spatially separated MZM pairs. Phase-shift error occurs when different qubits have different rotating period. Then, typical time of phase-shift is proportional to the difference between rotating periods of different qubits $a$ and $b$
\begin{eqnarray}\label{period}
t_{\textit{phase-shift}}\sim T_{a}-T_{b},
\end{eqnarray}
where $T$ is defined by Eq.~\ref{T}. For uniform system $H(r_i,r_j)=H(r_j,r_i)$, where $r_{i,j}$ denote by the MZM positions. In case of uniform system and uniform pertrubation all qubit states have the same rotation period, so no phase-shift errors occur. It means that phase-shift of the parity qubits occurs only in non-uniform system.

 Second, we deal with the bit-flip errors. This type of error correspond to the transitions of the state with empty fermion level to the state with full fermion level $|0\rangle \rightarrow |1\rangle$ and viceversa. This process is allowed now since PHC conjugation is broken. We can find amplitude of this process by calculating the matrix element
 \begin{eqnarray}
 \langle 0 |V_{as}|1\rangle\simeq
 (\langle \gamma_1|V_{as}|\gamma_1\rangle-\langle \gamma_2|V_{as}|\gamma_2\rangle)/\,2.
 \end{eqnarray}
 Once again, we omit factors $\langle \gamma_2|V_{as}|\gamma_1\rangle$ and $\langle \gamma_1|V_{as}|\gamma_2\rangle$ since this matrix elements are exponentially small if MZM are spatially separated. In case of uniform system this matrix element is zero. It means that bit-flips $|0\rangle \rightarrow |1\rangle$ can occur only in non-uniform system with typical time of the one bit-flip
 \begin{eqnarray}\label{bit_flip}
 t_{\textit{bit-flip}}\!\sim \!1/(\langle \gamma_1|V_{as}|\gamma_1\rangle-\langle \gamma_2|V_{as}|\gamma_2\rangle).
 \end{eqnarray}
\section{Origins of PHC asymmetry}\label{hhhz}
In this section we will study possible physical origin of PHC asymmetry in superconductors. We start with general form of PHC operator for the spinless case $\Xi=\tau_x K$. So, pertrubation that violates PHC symmetry $\Xi V_{as}(k)\Xi=+V_{as}(-k)$ can be written as $V_{as}(k)=\tau_x,\tau_y,\tau_0,\tau_z k$. For the spinful case PHC operator can be written as $\Xi=\sigma_y\tau_y K$. So, petrubation that violates PHC symmetry can be written as $V_{as}(k)=\sigma_i \tau_j, \tau_0, \sigma_i k, \tau_i k$. Here $\sigma_i$ acts in the spin space, $\tau_j$ acts in the particle-hole space, $i,j=x,y,z$, $\tau_0$ is the unity matrix, $K$ is the complex conjugation, $k$ is the momentum.

 Terms $\sigma_i\tau_x$ and $\sigma_i\tau_y$ corresponds to the superconducting order parameter with non-trivial spin structure. Terms $\sigma_i\tau_z$ corresponds to the asymmetric Zeeman field. So, PHC asymmetry can have either magnetic or superconducting origin, or its combination.
%1. \textit{Particle-hole asymmetric magnetic field due to particle-hole asymmetric g factor}. It has been argued in work Ref.~\onlinecite{Asano} based on the model Hamiltonian of topological insulator form Ref.~\onlinecite{zhang_model} that particle-hole asymmetric Zeeman splitting can be dominant in topological insulator based structures. Also, many experimental works on InAs quantum dots shows that electron and holes $g$-factor can be different~\cite{bennet_a,nakaoka_a}. So, if topological superconductor in a large Zeeman fields has particle-hole asymmetry in a $g$-factor then time for qubit operations can be limited by Eqs.~\ref{period} and ~\ref{bit_flip}. How we can eliminate this parasitic effect? First, we can use topological insulator-superconductor structures that do not require large magnetic fields. Second, we should use the same material for all qubits and use uniform magnetic field to provide the same period of qubit rotations and to eliminate bit-flip errors.
We want to note that in standard mean-field Fermi liquid theory such PHC asymmetric terms do not appear, since we cannot write them down in the standard second quantization form. Nevertheless, PHC asymmetry can arise in the Gutzwiller-projected mean-field theory, which is non-Fermi liquid theory that arrived from the Hubbard Hamiltonian~\cite{gutz_1,gutz_2,and}. So, we can assume that PHC asymmetry could manifest itself in the states with strong electron-electron correlations.

1. \textit{Intrinsic particle-hole asymmetry in high-temperature superconductors}. It has been argued by Hirsch that some high-temperature superconductors can have intrinsic PHC asymmetry due to electron-electron interactions~\cite{hirsch_polaron,hirsch_retarded,hirsch_not}. Recent experiments show that several high-temperature superconductors have intrinsic PHC asymmetry in the pseudogap state~\cite{pseudo_a} and in the Mott state~\cite{mott_a}. Also, PHC symmetry implies $\Xi\psi_E=\psi_{-E}$, so every state has its partner with opposite energy. It leads to the symmetric differential bias conductance near the Fermi level. However, experiments with cuprates show that this picture is not symmetric with respect to the Fermi level~\cite{exp_as1}, meaning that PHC symmetry is broken.

2. \textit{Particle-hole asymmetry in fractional quantum hall effect}. Usual fractional quantum hall effect have PHC symmetry and MZM can be realized as Moore-Read state with filling factor $\nu=5/2$~\cite{moore_read}. However, three body interaction causes Landau levels mixing and break PHC symmetry~\cite{halperin_break,fqhe_b1, fqhe_b2}. Such Landau level mixing removes the degeneracy between Pfaffian and its PHC anti-Pfaffian. This is the same that PHC asymmetric pertrubation $V_{as}$ do with MZM and its particle-hole conjugation. Recent experiments shows the presence of PHC asymmetry and Landau mixing in the high-quality samples of GaAs~\cite{fqhe_e1,fqhe_e2}.
\section{Discussion}
Previous studies of decoherence in Majorana qubits shows that stability of the MZM is not imply that TQC using MZM are stable. Indeed, tunneling of quasiparticles in the system do not affect MZM but it destroys coherence in the system~\cite{d1,d3}. In our case, we have surprisingly different result: TQC are stable versus PHC asymmetric pertrubation even if MZM are not stable.

If the system is uniform, then negative effects of the PHC asymmetry on decoherence of parity qubits are exponentially small, according to Eqs.~\ref{period},~\ref{bit_flip}. However, if we perform braiding operation using pairwise interactions~\cite{Akhmerov_anyon_braid} then the system becomes non-unform and decoherence effects occurs.

Recent works on Majorana qubits consider tunneling between MZM as one of the major source of decoherence~\cite{milestone}. Decoherence due to PHC asymmetry can be easily included to their results: we can use analogy between PHC asymmetry and coupling to the 'fake' MZM. In the formulas we just need to replace energy splitting caused by tunneling between separate MZM with first order correction to the energy caused by PHC asymmetric pertrubation.

In the Sec.~\ref{hhhz} it is shown that the PHC asymmetry can occur in the strongly correlated systems with a strong many-body interactions, such as fractional quantum Hall effect and cuprates. It means that systems with strong many-body interaction (for example, three-body interaction) may not be favorable for TQC. However, general description of the PHC asymmetry in such a system still lacks.

Simplest case of topological superconductor is $p$-wave superconductors. If PHC symmetry is exact, then this system is in $D$ symmetry class~\cite{schnyder}. In this class non-trivial $Z_2$ topological index can be defined, consequently, MZM can appear at a point defects in such a system~\cite{kane_defects}. PHC asymmetric pertrubation reduces symmetry class $D$ to the symmetry class $A$, where topological index for point defects is trivial, so there is no MZM ~\cite{kane_defects}. In our case it means that PHC asymmetric perturbation converts MZM to the single Dirac fermion with nonzero energy. However, even in this case quantum computations using Majorana qubits can be efficiently done. It arises interesting question: what properties of the system make it suitable for topological quantum computations?

\section{Conclusion}
In this work we investigated the influence of PHC asymmetric pertrubation on MZM based quantum computations. We found that PHC asymmetric pertrubation shift MZM from zero energy and add correction to the wavefunctions, which can be interpreted as an appearance of 'fake' MZM. We studied stability of MZM based parity qubits. We get that phase-shift and bit-flip errors occur due to PHC asymmetric pertrubation. We discussed possible origins of such asymmetry and argue that this asymmetry can arise in fractional quantum hall states with Landau mixing and high-temperature superconductors.

We acknowledge partial support by the Dynasty Foundation and
ICFPM (MMK), the Ministry of Education and Science of
the Russian Federation (Grant No. 14Y26.31.0007).

\end{document}